\documentclass[journal]{IEEEtran}
\usepackage{url}
\usepackage{natbib}

\begin{document}

\author{Russell Beale%
\thanks{R.Beale@bham.ac.uk\\
School of Computer Science, University of Birmingham, Edgbaston,\\ 
Birmingham, B15 2TT, UK\\
(c) 2025 IEEE. All rights reserved, including rights for text and
data mining, and training of artificial intelligence and similar technologies.}
}
 
% The paper headers
\markboth{IEEE Computer}{IEEE Computer}

\title{In Memoriam: The Academic Journal}

\maketitle
% As a general rule, do not put math, special symbols or citations
% in the abstract or keywords.
\begin{abstract}

In this piece we reflect on the life and influence of AJ, the academic journal, charting their history and contributions to science, discussing how their influence changed society and how, in death, they will be mourned for what they once stood for but for which, in the end, they had moved so far from that they will less missed than they might have been. 
\end{abstract}

% Note that keywords are not normally used for peerreview papers.

\section{Early Years}
Born in 1665 with the initial name "Journal des Sçavans", the Academic Journal (lovingly referred to as "AJ" from here onwards) toddled into the still active "Philosophical Transactions of the Royal Society" in the same year. AJ set a new approach to dissemination, being written by scholars and academics rather than journalists.  Early scientific journals embraced several approaches: some were run by individuals who exerted  control over the contents while others employed a group decision-making process, more closely aligned to modern peer review.  Early monographs proved primacy in discovery but did little to educate others or advance the field, and it took time for the dissemination of scientific information to the wider audience to gain traction. 

Initially parented by the learned societies, the teen years for AJ showed some focus on the world of work, providing professional bodies with their own specialist publications, notably the Lancet (1823), the British Medical Journal (1840), the Engineer (1856) and the Solicitor's Journal a year later in 1857.

In 1869 Nature was first published, and then Science (1880) published by the American Association for the Advancement of Science (AAAS).

\section{Early Adulthood}
 In the middle of the 20th Century,  AJ fully embraced  the concept of peer review, in which communities of like-minded academics checked and anlayzed the claims made in papers to ensure that they were, as far as possible, expressed clearly, and with enough detail to allow others to reproduce or replicate them.

Commercial influences rose to prominence just after the second world war, when publishing houses realised that both the raw material of content and the intellectual effort required to quality check it and shape it into publishable output would be provided, free of charge, by the academics, and that these same academics would pressure their institutions into paying large sums of money to be able to read what they had created.  As a commercial model, this is quite appealing --- take something that has free inputs and can be sold to a captive marketplace for substantial profits, and so Reed Elsevier, Springer Nature, Wiley-Blackwell, Taylor \& Francis, and SAGE came to dominate AJ's life with over 50\% of the market and 40\% profit margins.

\section{Middle Age}
Whilst Harvard reported on pressures to publish in 1939 \citep{harvard_university_report_1939}, most academic departments were expanding in the 1950's and 60's, in the US partly as a result of the "The Servicemen's Readjustment Act of 1944" (known as the GI Bill), which opened up higher education to most of the country, including about 7.7 million servicemen.  In 1947 half of all University students in the US were there because of the GI Bill.  Another expansionary push came from government funding to compete with the Soviet Union after Sputnik.

But as departmental growth slowed, we landed on the moon, and people kept working longer, so the rate of Professorial positions failed to keep pace with the number of people looking for such roles, and the era of competition arrived.  No doubt keen to capitalise on this, the publishers of AJ professed to offer a gatekeeping role of quality and pedigree; prestige on which to build a career.  This move from dissemination tool to measurement metric shifted AJ's philosophy completely, leading to a fateful meeting with a lareger than life character who came to dominate AJ's life.  This character was known as "PoP" - "Publish or Perish", a relationship that remained influential on AJ  until the end.

PoP insists that, for an academic to progress, they must amass a certain number of papers in a certain level of journal in order to be taken seriously by their peers and so either be hired, or progress through the ranks.  As domains became more specialised, arguments were made that within faculties, staff members could not be expected to be able to judge their colleagues' abilities to do research and so had to farm it out to the journals to vicariously do it for them, which was ironic since the journals just farmed it back to them, and charged them for the privilege. Academics being academics, they realised the rules of the game: if the metric was how any papers you published, then you needed to write in the MPU (the Minimum Publishable Unit), extracting two, three or more papers out of a single piece of work.  Of course, if you could also get a colleague to put your name on a paper they had written, then you would benefit too, and hence collaboration was born.  AJ embraced metrics: Impact Factor, h-index, i-index, total publications, number of collaborators --- all became measures of quality and professional esteem. The content of the work: well, let's consider the role of reviewing next.

Coupled with the growth of science in general, and technology, and professional content,  AJ became overwhelmed with material and decision times increased. Furthermore, the increased numbers of reviews to do, coupled with the increasing numbers of papers to write, meant that people had little time to replicate the experiments or results of others. Furthermore, doing so wasn't "new" and so less publishable, and so Publish or Perish push replication and validation down the academic agenda.  Reviewing became a major task, one not likely to advance a career, and so canny academics pushed this down onto graduate or undergraduate students, getting them to read and comment on work.  "Reviewer 2" as they are known by AJ and academics worldwide, are those reviewers who are not the clear and obvious first-choice academic to look at a paper, but the second opinion from whoever else can be persuaded to do it. Graduate students who know enough to have strong opinions but not enough to know good work from bad fill the ranks of "Reviewer 2" - most academic sin post have filled that role at some stage too in their past, no doubt.

\section{Children}
Arguably, these issues of lack of timeliness, increasingly arbitrariness of decisions, and lack of replication of results led to AJ considering their position in the world and becoming a parent, not so much out of choice, but out of necessity.  With the world-wide web in 1989 kickstarting a revolution in the way scientists found relevant papers by allowing electronic versions to be provided online, the rule of the printed versions of AJ's reign were doomed to the annals of history, with electronic publishing taking over.  AJ's first child, arXiv appeared in 1991, the first open access pre-print server, where anyone could upload a paper and, with minimal checks, it would be published online.  At its launch, arXiv focused on physics-related topics, specifically high-energy physics, condensed matter, and general relativity. Initially called the LANL preprint archive at xxx.lanl.gov, it expanded to include other disciplines like astronomy, mathematics, computer science, quantitative biology, and statistics. AJs child had a different philosophy to AJ --- rather than charging to read it, it would be free, and would allow academics to share their insights, data and research freely. Eschewing the rigours, or at least the trials, of peer-review, it championed open sharing of ideas over gatekeeping.  The influence of PoP impacted AJ's attitude towards their child: PoP still held sway in many institutions who didn't regard arXiv as premium enough to count, and so AJ remained current, paralleled but not usurped by their own offspring.   With Google launching in 1997, and Google Scholar in 2004, publishing had to be available online.  AJ produced a second offspring, when in 2003 PLOS (the Public Library of Science) was published, offering for free what AJ had previously charged for, and then  PLOS One in 2006 as an AJ megajournal.  These were more serious challengers to AJ, offering quality outputs that met with PoP's approval, causing AJ's commercial interests to be challenged.

\section{Health crisis}
In the early 21st Century, AJ entered a "Serials Crisis" in which AJs costs to institutions rose at a much greater rate than inflation, with AJ providing no more, and sometimes less, in return.  As a result of this, the Open Access (OA) movement came into being.  In 2002 the Creative Commons was founded in the US which provides a series of licences to facilitate fair use of freely provided materials, and Open Journal Systems releases software for the free management and publishing of scholarly journals.  Various other initiatives follow, but AJ found commercial ways to benefit from this: by allowing institutions to pay them not only for subscriptions to their journals but also to pay to publish articles, the AJs would then give free access to the articles to the academics.  It can best be explained by being a Schroedinger cost - free, and not free, at the same time.  The commercial interests sold the concept partly on the basis of PoP, and partly on the fact that you could have two forms of publication: Gold, which meant it was published immediately (by which they meant, after their usually quite long peer review process had completed) and pay handsomely for the privilege, or you could have Green option, in which they would delay it even longer, though it would be free.  And since publishing second isn't an option for most academics under PoP, then really only Gold was an option.  But hey, said AJ's backers, we're not forcing you.  Abetted by government subsidies to cover these costs, cash-strapped institutions signed up and a refreshed and re-invigorated AJ emerged into a new era as healthy as they had been before.

 \section{Early old age}
 Whilst in decent health, AJ felt  ongoing impacts form the lived so far most notably the long turnaround times. In fast moving fields such as Computer Science, the conference was rapidly becoming the publication venue of choice, owing to turnaround times for decisions in a few weeks rather than a few months, and because academics liked to take holidays in nice places too whilst calling it work.  Unsure how to respond, AJ and PoP conspired to allow authors to publish extended versions of conference paper in a journal - a win-win for all.  Not only did the academic get a holiday, they also got two papers for the price of one; AJ still got content, thanks to PoP and MPU.

 \section{The Covid-19 years}
 Covid-19 was a major boon for AJ  The pandemic led to a surge in research focused on Covid-19, which   boosted the prominence and impact of journals publishing this type of work. Their impact factor values increased, especially for lower-ranked journals, and some, like Critical Care Medicine, experienced increased citation activity per publication, this wasn't necessarily reflected in a sustained increase in total citations \citep{razavi_covid-19_2024}. The impact of COVID-19 on AJ publishing was multifaceted and may not have been uniformly felt across all journals \citep{uddin_impact_2023}.  The restrictions on in-person meetings travel and gatherings meant that academics could no longer attend their conferences, so benefitting AJ.

 \section{Sudden severe illness}
 In 2022, AJ seemed in great health, having weathered the conference issues and coming out of the pandemic stronger than ever, AJ looked to be set for a long retirement.  Little did they know that the contagion of their demise was already released and in the wild. ChatGPT had arrived.  Of wide public interest but little impact, 2022 saw large language models (LLMs) enter the mainstream for the first time - here was an artificial intelligence that was both fairly intelligent and not awfully painful to use.  Previously, AI systems had been like shouting down a telephone to get a robot to recognise which of four options you wanted, and failing miserably, yet here was a system that would chat you you and give you somewhat sensible answers.  As ChatGPT became more common, it started summarising conversations, and emails, and then documents. It started coding, and then doing research, and collating information on all sorts of things.  It seemed to have been trained mostly on fairy stories and creative fiction writing, because it constantly tried to create plausible sounding but completely made up information, giving rise to some hilarious mishaps when the well-intentioned but ill-informed used it for legal articles or, believe it or not, academic research.

 But in the depths of this wonderful, creative, supportive, occasionally hallucinogenic companion were the seeds of AJ's demise. In the past, PoP had pushed academics to trim the MPU down to the smallest size possible, to gain the maximum benefit from work as one could.  Indeed, even adding names to other's work with only a passing interest and sometimes an inability to even spell the methodological approach or pedagogic principle or reactive reagent would be overlooked as just something one had to do to get ahead in academic circles.  But fabricating results? That took time and effort, and writing all the surrounding guff to go with them took energy.  Sure, there had been some fraud - Andrew Wakefield famously fabricated evidence claiming a link between the MMR and vaccines, mostly because he stood to earn up to \$43 million a year from selling testing kits, so the effort seemed worthwhile. His discredited work caused a significant uptick in vaccine scepticism in the US \citep{motta_quantifying_2021} and multiple unnecessary deaths.  So it had been the case that, if you were going to fabricate results, you needed good financial reasons to do it,  But the advent of ChatGPt laid that to rest, Now you could create a decent scientific article with little to no effort --- some carefully engineered prompts to tame the creative excesses of the LLM, coupled with the comprehensive literature it could review, assimilate and represent in flowing, professional, understandable rose meant that large numbers of articles could be produced, based on highly limited, or even non-existent, or fabricated data --- and who would know?  AJ's pressures on reviewers hadn't diminished; Reviewer 2 was worn down to make minor comments only; no-one replicated results.  The flood of material overwhelmed the few remaining academics - editors, associate editors, diligent reviewers, who had been soldiering on to keep AJ going and the publishers wealthy.
 
And thus AJ entered the end-stage of life.  No longer could people rely on the content, because the cost of creating fake material was so low, and the benefits so high.  Now an academic could possibly publish half a dozen articles in a year, mostly because they could submit a hundred and hope a few got through.  AJ couldn't adapt - suddenly, their free raw material became mostly worthless, and the free processing they had relied on became worthless too, overwhelmed by quantity and plausible but uncheckable content.

The conference made a comeback.  With travel restrictions lifted, not only could the academic holidaying continue, but they could actually meet with fellow academics and quiz them on their findings to see if they were real: content had become checkable, and was king once again.  The actual exchange of information, findings and insights because important once again.

AJ died on 1st January 2026.  No flowers are expected, though AJ is still accepting paper submissions.  AJ is survived by colleagues PoP (who is reported to be a bit unwell, much othe delight of many academics), Reviewer 2, MPU, and offspring OA, arXiv and Plos, and a number of other AJ descendants.

\bibliographystyle{abbrvnat}
\renewcommand\refname{Mourners} 
\bibliography{references}

\end{document}